\documentstyle[12pt]{article}

\begin{document}

\setcounter{page}{1}

\pagestyle{plain} \vspace{1cm}
\begin{center}
\Large{\bf Cosmological dynamics of a non-minimally coupled bulk scalar field in DGP setup }\\
\small \vspace{1cm} {\bf Kourosh
Nozari\footnote{knozari@umz.ac.ir}}\quad and\quad {\bf Narges Rashidi\footnote{n.rashidi@umz.ac.ir}}\\
\vspace{0.5cm} Department of Physics, Islamic Azad University, Sari
Branch, Sari, Iran
\end{center}

\vspace{1cm}
\begin{abstract}
We consider cosmological dynamics of a canonical bulk scalar field,
which is coupled non-minimally to 5-dimensional Ricci scalar in a
DGP setup. We show that presence of this non-minimally coupled bulk
scalar field affects the jump conditions of the original DGP model
significantly. Within a superpotential approach, we perform some
numerical analysis of the model parameter space and consider
bulk-brane energy exchange in this setup. Also we show that the
normal, ghost-free branch of the DGP solutions in this case has the
potential to realize a self-consistent phantom-like behavior and
therefore explains late time acceleration of the universe in a
consistent way.\\
{\bf PACS}: 04.50.-h, \,98.80.Cq,\, 95.36.+x,\, 04.65.+e \\
{\bf Key Words}: Braneworld Cosmology, Induced Gravity, Bulk Scalar
Field, Superpotentials, Phantom-like Behavior.
\end{abstract}
\newpage

\section{Introduction}
According to the recent cosmological observations, our universe is
undergoing an accelerating phase of expansion and transition to the
accelerated phase has been occurred in the recent cosmological past
[1]. The simplest way to describe the accelerated expansion of the
universe is to adopt a cosmological constant. However, huge amount
of fine-tuning required for its magnitude and other theoretical
problems such as unknown origin and lake of dynamics make it
unfavorable for cosmologists [2]. So, to explain this remarkable
behavior of the universe, many theoretical approaches have been
proposed in recent years [3,4].

Other alternative approaches to accommodate dark energy are
modification of general relativity by considering additional spatial
dimensions [5-8]. Also teleparallel gravity provides a basis for
explanation of the late-time accelerated expansion [9]. In the
revolutionary braneworld viewpoint, our universe is a $3$-brane
embedded in an extra dimensional bulk. Standard matter and all
interactions are confined on the brane; only graviton and possibly
non-standard matter are free to probe the full bulk. One of the
various braneworld scenarios, is the model proposed by Dvali,
Gabadadze and Porrati (DGP). This setup is based on a modification
of the gravitational theory in an induced gravity perspective [7,8].
This induced gravity term in the model leads to deviations from the
standard 4-dimensional gravity over large distances. In this
braneworld scenario, the bulk is considered as empty except for a
cosmological constant and the matter fields on the brane are
considered as responsible for the evolution on the brane. Although
the DGP setup is successful to explain late-time acceleration of the
universe expansion in its self-accelerating branch, the model has
ghost instability in this branch [10]. Nevertheless, the normal,
non-self-accelerating branch is ghost-free and as has been shown,
has the potential to realize interesting cosmological implications
[11,12]. On the other hand, when a higher dimensional embedding
space exists, we are free to consider some bulk matter which can
certainly influence the cosmological evolution on the brane and can
be a major contributor to the dark energy. One of the particular
forms of bulk matter is a scalar field [13]. We have studied with
details the cosmological dynamics of \emph{minimally} coupled bulk
scalar field in the DGP setup recently [14,15]. But, since scalar
field can interact with other fields such as the gravitational
sector of the theory, in the spirit of braneworld scalar-tensor
theories, we can consider a non-minimal coupling (NMC) of the scalar
field with intrinsic (Ricci) curvature induced on the brane. We note
that generally the introduction of the NMC is not just a matter of
taste. The NMC is instead forced upon us in many situations of
physical and cosmological interest. There are compelling reasons to
include an explicit non-minimal coupling in the action. For
instance, non-minimal coupling arises at the quantum level when
quantum corrections to the scalar field theory are considered. Even
if for the classical, unperturbed theory this non-minimal coupling
vanishes, it is necessary for the renormalizability of the scalar
field theory in curved space. In most theories used to describe
inflationary scenarios, it turns out that a non-vanishing value of
the coupling constant cannot be avoided [16,17]. Generalization of
this scalar-tensor extension to braneworld setup with modified
induced gravity in the spirit of $f(R)$ theories has been considered
recently [18].

In the present work, we consider a scalar field which is
non-minimally coupled with intrinsic (Ricci) curvature in the bulk
of a DGP braneworld scenario. In section ${\bf 2}$ we derive the
Bulk-brane Einstein's equations and the scalar field's equation of
motion. Also, we find the jump conditions on the brane in the
presence of this bulk scalar field. In section ${\bf 3}$ we consider
the cosmology of this DGP-inspired model and find the Friedmann
equation for cosmological dynamics on the brane. In this section we
reduce the original partial differential field equations to an
ordinary differential equation. In section ${\bf 4}$, we use the
superpotential method to perform some numerical analysis on the
parameter space of the model and discuss on bulk-brane energy
exchange. We show that, since we expect the energy leaks of the
brane with expansion of the universe, there is some constraints on
parameter space of this scenario. In section ${\bf 5}$, we study the
late time behavior of this model. We show the normal branch of this
DGP-inspired model, in the presence of non-minimally coupled bulk
scalar field, can explain the late-time cosmic acceleration.

\section{The Setup}
The 5-dimensional action for a DGP-inspired braneworld model in the
presence of a non-minimally coupled scalar field in the bulk can be
written as follows
\begin{equation}
S=\int_{B}d^{5}x\sqrt{-g}\Big\{\frac{1}{2\kappa_{5}^{2}}\,
(1-\xi\phi^{2})\,{\cal{R}}-\frac{1}{2}(\nabla\phi)^{2}-V(\phi)\Big\}
+\int_{b}d^{4}x\sqrt{-h}\Big(\frac{1}{2\kappa_{4}^{2}}R+\frac{1}{2\kappa_{5}^{2}}[K]+{\cal{L}}_{b}(\phi)
+{\cal{L}}_{b}^{(m)}\Big)\,,
\end{equation}
where $g_{AB}$ is the bulk metric and $h_{AB}$ is the induced metric
on the brane. They are related by $h_{AB}=g_{AB}-n_{A}n_{B}$, where
$n_{A}$ is the unit vector normal to the 3-brane and $A, B$ are the
five dimensional indices. The Gibbons-Hawking boundary term is
included via jump of the trace of the extrinsic curvature $[K]$\, in
the brane action.\, Also, $\kappa_{5}^2=\frac{8\pi}{M_{5}^3}$, where
$M_{5}$ is the fundamental 5-dimensional Planck mass. The brane
Lagrangian ${\cal{L}}_{b}(\phi)$ includes the Standard Model fields
which are confined to the brane and depends on the bulk scalar
field. It should be noticed that ordinary matter (such as dust
matter or perfect fluid) described by ${\cal{L}}_{b}^{(m)}$, is
confined to the brane located at $y=0$. In this action, the
non-minimal coupling of the bulk scalar field and the 5-dimensional
Ricci scalar is represented by the term $\xi\phi^{2}\,{\cal{R}}$. We
note that, the indices `` B " and `` b " in integral refer to ``
Bulk " and `` brane " respectively.\\

The Bulk-brane Einstein's equations calculated from action (1) are
given by
\begin{equation}
\frac{1}{\kappa_{5}^{2}}\Big(1-\xi\phi^{2}\Big)\Big({\cal{R}}_{AB}-\frac{1}{2}g_{AB}{\cal{R}}\Big)
+\frac{1}{\kappa_{4}^{2}}\Big(R_{AB}-\frac{1}{2}h_{AB}R\Big)\delta(y)=T_{AB}^{(B)}+T_{AB}^{(b)}\delta(y),
\end{equation}
where $T_{AB}^{(B)}$ and $T_{AB}^{(b)}$ are the bulk and brane
energy momentum tensor respectively. Also, $\delta(y)$ is the Dirac
delta function with support on the brane which we assume to be
located at $y=0$, where $y$ is the coordinate of the extra
dimension. In our setup, the energy momentum of the bulk is given by
following expression
$$
T_{AB}^{(B)}=\nabla_{A}\phi\,\nabla_{B}\phi-\frac{1}{2}\nabla^{K}\phi\,\nabla_{K}\phi
\,g_{AB}-V(\phi)\,g_{AB}
$$
\begin{equation}
+2\,\xi\,\nabla_{A}\phi\,\nabla_{B}\phi-2\,\xi\,\nabla^{K}\phi\,\nabla_{K}\phi
\,g_{AB}+2\,\xi\,\phi\,\nabla_{A}\nabla_{B}\phi-2\,\xi\,\phi\,\nabla^{K}\nabla_{K}\phi\,
g_{AB}.
\end{equation}
This energy momentum tensor leads to the following bulk energy
density and pressure
$$
\rho^{(B)}=2n^{2}\xi\Bigg(\phi'^{2}+\phi\phi''\Bigg)+6n^{2}\xi\Bigg(\frac{a'}{a}\Bigg)\phi\phi'
-6\xi\Bigg(\frac{\dot{a}}{a}\Bigg)\phi\dot{\phi}
$$
\begin{equation}
+\frac{1}{2}\dot{\phi}^{2}+\frac{n}{2}\phi'^{2}+n^{2}V(\phi),
\end{equation}

$$
p^{(B)}=-\frac{k}{a^{2}}\xi\phi^{2}+\frac{2\xi}{n^{2}}\Bigg[\dot{\phi}^{2}+\phi\ddot{\phi}+\bigg(2\frac{\dot{a}}{a}
-\frac{\dot{n}}{n}\bigg)\phi\dot{\phi}\Bigg]
-2\xi\Bigg[\phi'^{2}+\phi\phi''+\bigg(2\frac{a'}{a}+\frac{n'}{n}\bigg)\phi\phi'\Bigg]
$$
\begin{equation}
+\frac{1}{2n^{2}}\dot{\phi}^{2}-\frac{1}{2}\phi'-V(\phi).
\end{equation}
We note that, the first three terms in equations (4) and (5), show
the effect of non-minimal coupling in the bulk and if we set
$\xi=0$, these equations simplify to the usual results for minimal
case. $T_{AB}^{(b)}$, the total energy momentum of the brane, is
defined as
\begin{equation}
T_{AB}^{(b)}=T_{AB}^{(m)}+T_{AB}^{(\phi)},
\end{equation}
where $T_{AB}^{(\phi)}$ and $T_{AB}^{(m)}$ are the energy momentum
tensor corresponding to ${\cal{L}}_{b}(\phi)$ and
${\cal{L}}_{b}^{(m)}$ respectively. The scalar field's equation of
motion given by
\begin{equation}
\nabla^{2}\phi-n^{2}\frac{dV}{d\phi}+\frac{n^{2}}{\kappa_{5}^{2}}\xi\phi{\cal{R}}
+n^{2}\frac{\sqrt{-h}}{\sqrt{-g}}\frac{d{\cal{L}}_{b}(\phi)}{d\phi}=0.
\end{equation}
The action (1) implies the following jump conditions
\begin{equation}
\Big[N^{A}\nabla_{A}\phi\Big]=\frac{\delta
{\cal{L}}_{b}(\phi)}{\delta \phi}\,,
\end{equation}
\begin{equation}
\Big[K_{AB}-Kh_{AB}\Big]=-\kappa_{4}^{2}T_{AB}^{(b)}.
\end{equation}

To formulate cosmological dynamics on the brane, we assume the
following line element
\begin{equation}
ds^{2}=g_{AB}dx^{A}dx^{B}=-n^{2}(y,t)dt^{2}+a^{2}(y,t)\gamma_{ij}dx^{i}dx^{j}+b^{2}(y,t)dy^{2},
\end{equation}
where $\gamma_{ij}$ is a maximally symmetric 3-dimensional metric
defined as $\gamma_{ij}=\delta_{ij}+k\frac{x_{i}x_{j}}{1-kr^{2}}$
where $k=-1,0,+1$ parameterizes the spatial curvature and
$r^{2}=x_{i}x^{i}$.

Since here we consider homogeneous and isotropic geometries inside
the brane, $T_{B}^{A(b)}$ can be expressed quite generally in the
following form
\begin{equation}
T_{B}^{A(b)}=\frac{1}{b}diag\Big(-\rho^{(b)},\,\,\,p^{(b)},\,\,\,
p^{(b)},\,\,\,p^{(b)},\,\,\,0\Big)\,,
\end{equation}
where $\rho^{(b)}=\rho^{(m)}+\rho^{(\phi)}$. Note that here
$\rho^{(\phi)}$ is the energy density corresponding to
$T_{AB}^{(\phi)}$ and $\rho^{(m)}$ is the energy density
corresponding to $T_{AB}^{(m)}$. The extrinsic curvature tensor in
the background metric (10) is given by
\begin{equation}
K_{B}^{A}=diag\Big(\frac{n'}{nb}\,,\,\,\frac{a'}{ab}\delta_{j}^{i}\,,\,\,0\Big).
\end{equation}

By using the metric ansatz (10) and adopting the Gaussian normal
coordinates $(b(y,t)=1)$, we obtain the equations of motion in the
following form
$$
\frac{3}{\kappa_{5}^{2}}\Bigg(1-\xi\phi^{2}\Bigg)\Bigg(\frac{\dot{a}^{2}}{a^{2}}-\frac{n^{2}a'^{2}}{a^{2}}
-\frac{n^{2}a''}{a}+\frac{n^{2}k}{a^{2}}\Bigg)
+\frac{3}{\kappa_{4}^{2}}\Bigg(\frac{\dot{a}^{2}}{a^{2}}+\frac{n^{2}k}{a^{2}}\Bigg)\delta(y)
=\frac{1}{2}\dot{\phi}^{2}+\frac{n^{2}}{2}\phi'^{2}
+n^{2}V(\phi)
$$
\begin{equation}
-2n^{2}\xi\Bigg(\phi'^{2}+\phi\phi''\Bigg)-6n^{2}\xi\Bigg(\frac{a'}{a}\Bigg)\phi\phi'
+6\xi\Bigg(\frac{\dot{a}}{a}\Bigg)\phi\dot{\phi}
+n^{2}\rho^{(b)}\delta(y),
\end{equation}\\
$$
\gamma_{ij}\frac{a^{2}}{\kappa_{5}^{2}}\Bigg(1-\xi\phi^{2}\Bigg)\left[\left(\frac{a'^{2}}{a^{2}}-\frac{\dot{a}^{2}}{n^{2}a^{2}}
-\frac{k}{a^{2}}\right)
+2\left(\frac{a''}{a}+\frac{n'a'}{na}-\frac{\ddot{a}}{n^{2}a}+\frac{\dot{n}\dot{a}}{n^{3}a}+\frac{n''}{2n}\right)\right]
$$
$$
+\frac{\gamma_{ij}}{\kappa_{4}^{2}}\left[2\left(\frac{\dot{n}\dot{a}}{n^{3}a}-\frac{\ddot{a}}{n^{2}a}\right)
-\left(\frac{\dot{a}^{2}}{n^{2}a^{2}}+\frac{k}{a^{2}}\right)\right]\delta(y)=k\gamma_{ij}\Bigg(1-\xi\phi^{2}\Bigg)
+a^{2}\gamma_{ij}\Bigg[\frac{1}{2n^{2}}\dot{\phi}^{2}-\frac{1}{2}\phi'^{2}-V(\phi)\Bigg]
$$
\begin{equation}
+2\xi\gamma_{ij}\Bigg\{\frac{a^{2}}{n^{2}}\Bigg[-\Big(\dot{\phi}^{2}+\phi\ddot{\phi}\Big)
-\bigg(2\frac{\dot{a}}{a}-\frac{\dot{n}}{n}\bigg)\phi\dot{\phi}\Bigg]+
a^{2}\Bigg[\bigg(\phi'^{2}+\phi\phi''\bigg)+\bigg(2\frac{a'}{a}+\frac{n'}{n}\bigg)\phi\phi'\Bigg]\Bigg\}
-a^{2}\gamma_{ij}p^{(b)}\delta(y),
\end{equation}\\
\begin{equation}
\frac{3}{\kappa_{5}^{2}}\Bigg(1-\xi\phi^{2}\Bigg)\Bigg(\frac{n'}{n}\frac{\dot{a}}{a}-\frac{\dot{a}'}{a}\Bigg)
=-2\xi\Bigg(\phi'\dot{\phi}+\phi\dot{\phi}'\Bigg)
+2\xi\Bigg(\frac{n'}{n}\Bigg)\phi\dot{\phi}+\dot{\phi}\phi',
\end{equation}\\
$$
\frac{3}{\kappa_{5}^{2}}\Bigg(\frac{a'^{2}}{a^{2}}-\frac{\dot{a}^{2}}{n^{2}a^{2}}-\frac{k}{a^{2}}+\frac{n'a'}{na}
+\frac{\dot{n}\dot{a}}{n^{3}a}
-\frac{\ddot{a}}{n^{2}a}\Bigg)=\frac{1}{2}\phi'^{2}+\frac{1}{2n^{2}}\dot{\phi}^{2}-V(\phi)
+2\xi\Bigg(3\frac{a}{a'}-\frac{n'}{n}\Bigg)\phi\phi'
$$
\begin{equation}
-\frac{1}{n^{2}}\Bigg[2\xi\bigg(\dot{\phi}^{2}+\phi\ddot{\phi}\bigg)+2\xi\bigg(3\frac{\dot{a}}{a}
-\frac{\dot{n}}{n}\bigg)\phi\dot{\phi}\Bigg],
\end{equation}\\
where a prime marks differentiation with respect to $y$ and a dot
denotes differentiation with respect to $t$. Equations (13)-(16) are
$(0,0)$, $(i,j)$, $(0,5)$ and $(5,5)$ components of Einstein field
equations respectively. Also, the scalar field evolution equation is
given by
\begin{equation}
\ddot{\phi}+\bigg(3\frac{\dot{a}}{a}-\frac{\dot{n}}{n}\bigg)\dot{\phi}
-n^{2}\Bigg[\phi''+\bigg(\frac{n'}{n}+3\frac{a'}{a}\bigg)\phi'\Bigg]
+n^{2}\frac{dV}{d\phi}+\frac{n^{2}}{\kappa_{5}^{2}}\xi\phi{\cal{R}}
+n^{2}\frac{\sqrt{-h}}{\sqrt{-g}}\frac{d{\cal{L}}_{b}(\phi)}{d\phi}\delta(y)=0,
\end{equation}
where the bulk Ricci scalar is defined as follows
\begin{equation}
{\cal{R}}=3\frac{k}{a^{2}}+\frac{1}{n^{2}}\Bigg(6\frac{\ddot{a}}{a}+6\frac{\dot{a}^{2}}{a^{2}}
-6\frac{\dot{a}\dot{n}}{an}\Bigg)-6\frac{a''}{a}-2\frac{n''}{n}-6\frac{a'^{2}}{a^{2}}-6\frac{a'n'}{an}.
\end{equation}

As we know, in order to have a well-defined geometry, the metric is
required to be continuous across the brane localized in $y=0$.
However, its derivatives with respect to $y$ can be discontinuous in
$y=0$. There are some terms in the Einstein tensor components that
are second derivative of the metric ($a''$, $\phi''$, $n''$). One
can decompose these terms as [19]
\begin{equation}
a''=\hat{a}''+[a']\,\delta(y),
\end{equation}
\begin{equation}
n''=\hat{n}''+[n']\,\delta(y),
\end{equation}
and
\begin{equation}
a''=\hat{a}''+[a']\,\delta(y),
\end{equation}
where $\hat{A}$ shows the non-distributional part of the quantity
$A$ and $[A] \equiv A(0^{+}) - A(0^{-})$ shows the jump of this
quantity across $y=0$. With this decomposition, one can equate those
terms containing a Dirac delta function in the Einstein tensor with
the distributional components in the stress-energy tensor. This
matching leads to the following relations for the jump conditions
$$
\frac{[a']}{a_{0}}=\frac{4\xi^{2}\phi_{0}^{2}h_{0}}{n_{0}^{2}A_{0}}+\frac{12\xi^{2}\phi_{0}^{2}k}{a_{0}^{2}A_{0}}
+\frac{2\kappa_{5}^{2}\xi\phi_{0} }{A_{0}}\frac{\delta
{\cal{L}}(\phi)}{\delta\phi}+\frac{3\kappa_{5}^{2}l_{0}}{n_{0}^{2}\kappa_{4}^{2}A_{0}}-\frac{\kappa_{5}^{2}}{A_{0}}\rho^{(b)}
-\frac{8\kappa_{5}^{2}\xi^{2}\phi_{0}^{2}}{(1-\xi\phi_{0}^{2})A_{0}}\Big(\rho^{(b)}+p^{(b)}\Big)
$$
\begin{equation}
+\frac{16\kappa_{5}^{2}\xi^{2}\phi_{0}^{2}}{n_{0}^{2}\kappa_{4}^{2}(1-\xi\phi_{0}^{2})A_{0}}\left(l_{0}+m_{0}n_{0}^{2}\right)\,,
\end{equation}

$$
\frac{[\phi']}{\phi_{0}}=\frac{6\xi
h_{0}}{n_{0}^{2}\kappa_{5}^{2}A_{0}}(1-\xi\phi_{0}^{2})+\frac{18\xi
k}{\kappa_{5}^{2}a_{0}^{2}A_{0}}(1-\xi\phi_{0}^{2})+\frac{3(1-\xi\phi_{0}^{2})}{\phi_{0}
A_{0}}\frac{\delta
{\cal{L}}(\phi)}{\delta\phi}-\frac{24\xi}{n_{0}^{2}\kappa_{4}^{2}A_{0}}\left(l_{0}-m_{0}n_{0}^{2}\right)
$$
\begin{equation}
-\frac{4\xi}{A_{0}}\left(3p^{(b)}-\rho^{(b)}\right)\,,
\end{equation}

and
$$
\frac{[n']}{n_{0}}=\frac{2\kappa_{5}^{2}\xi\phi_{0}
}{A_{0}}\frac{\delta
{\cal{L}}(\phi)}{\delta\phi}+\frac{2\kappa_{5}^{2}}{A_{0}}\rho^{(b)}+\frac{3\kappa_{5}^{2}}{A_{0}}p
+\frac{12\xi^{2}\phi_{0}^{2}k}{a_{0}^{2}A_{0}}+\frac{4\xi^{2}\phi_{0}^{2}h_{0}}{n_{0}^{2}A_{0}}
-\frac{3\kappa_{5}^{2}l_{0}}{n_{0}^{2}\kappa_{4}^{2}A_{0}}
-\frac{6\kappa_{5}^{2}m_{0}}{\kappa_{4}^{2}A_{0}}
$$
\begin{equation}
+\frac{24\kappa_{5}^{2}\xi^{2}\phi_{0}^{2}}{(1-\xi\phi_{0}^{2})A_{0}}(\rho^{(b)}+p^{(b)})
-\frac{48\kappa_{5}^{2}\xi^{2}\phi_{0}^{2}}{n_{0}^{2}\kappa_{4}^{2}(1-\xi\phi_{0}^{2})A_{0}}(l_{0}+m_{0}n_{0}^{2})\,,
\end{equation}
where a prime marks differentiation with respect to $y$, a dot
denotes differentiation with respect to $t$ and the subscript $0$
marks quantities that are calculated at $y=0$ (on the brane). Also
we have defined the following parameters
\begin{equation}
l=\frac{\dot{a}^{2}}{a^{2}}+\frac{n^{2}k}{a^{2}}\,,
\end{equation}
\begin{equation}
m=\frac{\dot{n}\dot{a}}{n^{3}a}-\frac{\ddot{a}}{n^{2}a}\,,
\end{equation}
\begin{equation}
h=6\frac{\ddot{a}}{a}+6\frac{\dot{a}^{2}}{a^{2}}-6\frac{\dot{a}\dot{n}}{an}\,,
\end{equation}
\begin{equation}
A=3-3\xi\phi^{2}+32\xi^{2}\phi^{2}\,.
\end{equation}

Assuming $Z_{2}$-symmetry about the brane for simplicity, the
junction conditions (22)-(24) can be used to compute $a'$, $n'$ and
$\phi'$ on two sides of the brane. It should be noticed that if we
consider the case with $\xi=0$ (the minimal coupling between the
scalar field and the bulk Ricci scalar), the jump conditions
(22)-(24) simplify to the jump conditions achieved in Ref. [15].\\

If we take the jump of the component (0,5) of Einstein equations and
use the jump conditions, we shall achieve the energy conservation
equation. In our setup this equation is a complicated expression as
follows
$$
\dot{\rho}^{(b)}+3\frac{\dot{a}_{0}}{a_{0}}(\rho^{(b)}+p^{(b)})=
-\frac{A_{0}}{3(1-\xi\phi^{2})_{y=0}}\Bigg\{\frac{96\xi^{2}\phi^{2}}{A}\left(\frac{\dot{a}}{a}\right)(\rho^{(b)}+p^{(b)})
+\frac{48\xi^{2}\phi\dot{\phi}}{A}(\rho^{(b)}+p^{(b)})+\frac{{\cal{X}}\left(\frac{\dot{a}}{a}\right)}{3\kappa_{4}^{2}A}h
$$
$$
-\frac{2\kappa_{5}^{2}\xi\phi\dot{\phi}}{A}(2\rho^{(b)}+3p^{(b)})+\frac{\dot{\phi}}{A}{\cal{A}}(\rho^{(b)}-3p^{(b)})
-36(1-\xi\phi^{2})\frac{\xi^{2}\phi\dot{\phi}}{\kappa_{5}^{2}A}h
+\frac{8\xi^{3}\phi^{2}\dot{\phi}}{\kappa_{5}^{2}A}\left[3\dot{\phi}-\kappa_{5}^{2}\phi\right]h
$$
$$
+6(1-\xi\phi^{2})\left[\frac{\dot{\phi}}{\kappa_{5}^{2}A}\right]{\cal{B}}h
+9(1-\xi\phi^{2})\left[\frac{2H+\frac{\dot{A}}{A}}{\kappa_{4}^{2}A}\right]\left(\frac{\dot{a}}{a}\right)^{2}
-\frac{48\xi^{2}\phi^{2}}{\kappa_{4}^{2}(1-\xi\phi^{2})A}\left[2\xi\phi\dot{\phi}
-\kappa_{5}^{2}\right]\left(\frac{\dot{a}}{a}\right)^{2}
$$
$$
+\frac{6\xi\phi\dot{\phi}}{\kappa_{5}^{2}A}{\cal{D}}H^{2}-\frac{18m}{\kappa_{4}^{2}A}{\cal{E}}
\left(\frac{\dot{a}}{a}\right)
-\frac{48\xi^{2}\phi^{2}}{\kappa_{4}^{2}(1-\xi\phi^{2})A}{\cal{K}}m-\frac{12\xi\phi\dot{\phi}}{\kappa_{4}^{2}A}{\cal{J}}m
-12(1-\xi\phi^{2})\frac{\xi^{2}\phi}{\kappa_{5}^{2}A}(\phi-\dot{\phi})\dot{h}
$$
$$
-\frac{48\xi^{2}\phi}{\kappa_{4}^{2}A}(\phi-\dot{\phi})\dot{m}
+3(1-\xi\phi^{2})\frac{{\cal{S}}}{A}+\frac{2\xi\dot{\phi}{\cal{W}}}{A}g
+\frac{8\xi^{2}\phi}{A}\left[3\phi(\dot{\rho}^{(b)}+\dot{p}^{(b)})+\dot{\phi}(3\dot{p}-\dot{\rho}^{(b)})\right]
$$
\begin{equation}
+\left[-3(1-\xi\phi^{2})\rho^{(b)}
-24\xi^{2}\phi^{2}(\rho^{(b)}+p^{(b)})\right]\frac{\dot{A}}{A}
\Bigg\}_{y=0}\equiv \Psi\,,
\end{equation}
where we have defined the following parameter
\begin{equation}
{\cal{A}}=8\xi^{2}\phi+4\xi\phi+\frac{8\xi^{2}\phi\dot{A}}{A}-8\xi^{2}\dot{\phi}\,,
\end{equation}
\begin{equation}
{\cal{B}}=-2\xi^{2}\dot{\phi}+\frac{2\xi^{2}\phi\dot{A}}{A}+\frac{2\xi^{2}\phi^{2}\dot{A}}{\dot{\phi}A}+\xi\phi\,,
\end{equation}
\begin{equation}
{\cal{D}}=-\frac{32\xi\phi}{\dot{\phi}}H-8\xi+\frac{8\xi\phi\dot{A}}{\dot{\phi}A}-4+\frac{8\xi\dot{\phi}}{\phi}
-\frac{8\dot{A}}{A}+\kappa_{5}^{2}\,,
\end{equation}
\begin{equation}
{\cal{E}}=1-\xi\phi^{2}+\frac{32}{3}\xi^{2}\phi^{2}\,,
\end{equation}
\begin{equation}
{\cal{K}}=2\xi\phi\dot{\phi}+\kappa_{5}^{2}\,,
\end{equation}
\begin{equation}
{\cal{J}}=12\xi-4\xi\frac{\phi\dot{A}}{\dot{\phi}A}-2+4\xi\frac{\dot{\phi}}{\phi}-4\xi\frac{\dot{A}}{A}-\kappa_{5}^{2}\,,
\end{equation}
\begin{equation}
{\cal{X}}=-9+9\xi\phi^{2}-48\xi^{2}\phi^{2}+48\xi^{2}\phi\dot{\phi}\,,
\end{equation}
\begin{equation}
{\cal{S}}=2\xi\phi\dot{g}+4\xi\dot{\phi}g-2\xi\phi
g\frac{\dot{A}}{A}-\dot{\phi}g+2\xi\dot{\phi}\dot{g}\,,
\end{equation}
\begin{equation}
{\cal{W}}=-6\xi\phi-3\frac{\dot{A}}{A}+2\kappa_{5}^{2}\phi^{2}\,,
\end{equation}
and
\begin{equation}
g=\frac{\delta {\cal{L}}_{b}(\phi)}{\delta\phi}\,.
\end{equation}
Also the subscript $0$ indicates the parameter on the brane. The
right hand side of the conservation equation is non-zero and this
means that the energy on the brane is not conserved, i.e. there may
be leakage of energy-momentum from the brane or suction onto the
brane. In this respect, for negative values of $\Psi$,
energy-momentum leaks off the brane and for positive $\Psi$
energy-momentum flows from the bulk onto the brane.

\section{DGP braneworld cosmology with a non-minimally coupled bulk scalar field}

Now we follow Ref. [20] to obtain a special class of solutions for a
DGP braneworld cosmology with a non-minimally coupled bulk scalar
field. In this regard, we introduce the quantity ${\cal{F}}$ as a
function of $t$ and $y$ as follows
\begin{equation}
{\cal{F}}(t,y)=\left(\frac{a'}{ab}\right)^{2}-\left(\frac{\dot{a}}{an}\right)^{2}.
\end{equation}
So, we can rewrite the components (0,0) and (5,5) of the Einstein's
field equations in the bulk as follows
\begin{equation}
G_{0}^{0}-\frac{\dot{a}}{a'}G_{5}^{0}=(1-\xi\phi^{2})\frac{3}{2a^{3}a'}\,\partial_{y}(a^{4}{\cal{F}})
\end{equation}
\begin{equation}
G_{5}^{5}-\frac{a'}{\dot{a}}G_{0}^{5}=(1-\xi\phi^{2})\frac{3}{2a^{3}\dot{a}}\,\partial_{t}(a^{4}{\cal{F}})
\end{equation}
where $G_{AB}$ is defined as
\begin{equation}
G_{AB}=\bigg(1-\xi\phi^{2}\bigg)\bigg({\cal{R}}_{AB}-\frac{1}{2}g_{AB}{\cal{R}}\bigg)
\end{equation}

In the presence of the bulk scalar field, the left hand sides of
these two equations are not the same. But for special class of
solutions with $\phi=\phi(a)$, they are equivalent and in this case
${\cal{F}}={\cal{F}}(a)$. In this situation, both (41) and (42)
reduce to
$$
6\left(1-\xi\phi^{2}\right){\cal{F}}+\frac{3}{2}\left(1-\xi\phi^{2}\right)\left(a\frac{d
{\cal{F}}}{d
a}\right)+\frac{\kappa_{5}^{2}}{2}{\cal{F}}\left(a\frac{d \phi}{d
a}\right)^{2}-2\kappa_{5}^{2}\xi {\cal{F}}\left(a\frac{d \phi}{d
a}\right)^{2}-2\kappa_{5}^{2}\xi {\cal{F}}\phi\,
a^{2}\left(\frac{d^{2}\phi}{d a^{2}}\right)
$$
\begin{equation}
-6\kappa_{5}^{2}\xi\phi {\cal{F}}\left(a\frac{d \phi}{d
a}\right)-\kappa_{5}^{2}\xi\phi\left(a\frac{d {\cal{F}}}{d
a}\right)\left(a\frac{d \phi}{d
a}\right)-\frac{2\kappa_{5}^{2}\xi\phi}{a}{\cal{F}}+\kappa_{5}^{2}V(\phi)=0.
\end{equation}

We choose a Gaussian normal coordinate system so that
$b^{2}(y,t)=1$. Also we assume that $t$ as a proper cosmological
time on the brane has scaled so that $n_{0}=1$. By adopting a
$Z_{2}$ symmetry across the brane, equations (13) and (16) yield the
following generalization of the Friedmann equation for cosmological
dynamics on the DGP brane
$$
3\kappa_{5}^{2}(\xi\phi_{0}^{2}-1)H^{2}+24\xi^{2}\phi_{0}^{2}\kappa_{4}^{2}(\xi\phi_{0}^{2}-1)H^{2}
+8\xi^{2}\phi_{0}^{2}\kappa_{5}^{2}\kappa_{4}^{2}(\rho^{(b)}+p^{(b)})
-16\xi^{2}\phi_{0}^{2}\kappa_{5}^{2}(m_{0}+H^{2})
$$
\begin{equation}
+\rho^{(b)}\kappa_{5}^{2}\kappa_{4}^{2}(1-\xi\phi_{0}^{2})+2\xi\phi_{0}\kappa_{4}^{2}\kappa_{5}^{2}(\xi\phi_{0}^{2}-1)g_{0}
+24\xi^{2}\phi_{0}^{2}\kappa_{4}^{2}(1-\xi\phi_{0}^{2})m_{0} =\pm
2\kappa_{4}^{2}(1-\xi\phi_{0}^{2})A\sqrt{H^{2}-{\cal{F}}_{0}}
\end{equation}
where $H=\left(\frac{\dot{a}}{a}\right)_{y=0}$ and $\pm$ refers to
two branches of the DGP model. We have solved this equation for
$H^{2}$ and the result is presented in {\bf Appendix A}. If we
consider minimal coupling of the scalar field and 5D Ricci scalar
($\xi=0$), we reach the following Friedmann equation on the brane
which was obtained in [15]
\begin{equation}
H^{2}=\frac{1}{3}\kappa_{4}^{2}\rho^{(b)}+
\frac{2\kappa_{4}^{4}}{\kappa_{5}^{4}}\pm
\frac{2\kappa_{4}^{2}}{\kappa_{5}^{2}}\sqrt{\frac{\kappa_{4}^{4}}
{\kappa_{5}^{4}}+\frac{1}{3}\kappa_{4}^{2}\rho^{(b)}+{\cal{F}}_{0}}.
\end{equation}
We note that from now on we consider only the normal, ghost-free
branch of the Friedmann equation. Since we consider $\phi=\phi(a)$,
the scalar field equation (7) reduces to
\begin{equation}
\nabla^{2}\phi=a^{2}\Big(\frac{d^{2}\phi}{da^{2}}+\frac{1}{a}\frac{d\phi}{da}\Big){\cal{F}}
+\Big[\frac{a(G^{0}_{\,0}+G_{\,5}^{5})}{3(1-\xi\phi^{2})}\Big]\frac{d\phi}{da}=-\frac{1}{\kappa_{5}^{2}}\xi\phi
{\cal{R}}+\frac{d
V}{d\phi}-\delta(y)\frac{\sqrt{-h}}{\sqrt{-g}}\frac{\delta
{\cal{L}}_{b}(\phi)}{\delta\phi}.
\end{equation}

Now, by substituting equations $G_{AB}=\kappa_{5}^2 T_{AB}$ and (44)
into (47), we find
$$\frac{2a}{3(1-\xi\phi^{2})}\Bigg\{6\bigg(1-\xi\phi^{2}\bigg){\cal{F}}
+\frac{3}{2}\bigg(1-\xi\phi^{2}\bigg)\bigg(a\frac{d {\cal{F}}}{d
a}\bigg)+\frac{\kappa_{5}^{2}}{2}{\cal{F}}\bigg(a\frac{d \phi}{d
a}\bigg)^{2}-2\kappa_{5}^{2}\xi {\cal{F}}\phi\,
a^{2}\bigg(\frac{d^{2}\phi}{d
a^{2}}\bigg)-\frac{2\kappa_{5}^{2}\xi\phi}{a}{\cal{F}}
$$
$$
-2\kappa_{5}^{2}\xi {\cal{F}}\bigg(a\frac{d \phi}{d
a}\bigg)^{2}-6\kappa_{5}^{2}\xi\phi {\cal{F}}\bigg(a\frac{d \phi}{d
a}\bigg)-\kappa_{5}^{2}\xi\phi\bigg(a\frac{d {\cal{F}}}{d
a}\bigg)\bigg(a\frac{d \phi}{d a}\bigg)+2\kappa_{5}^{2}\xi\phi
\bigg(a''+\frac{n'a'}{a}-\frac{\ddot{a}}{n^{2}}\bigg)
\bigg(\frac{d\phi}{da}\bigg)\Bigg\}\frac{d\phi}{da}$$
\begin{equation}
+{\cal{F}}\bigg(a\frac{d}{da}\bigg)^{2}\phi+\frac{1}{\kappa_{5}^{2}}\xi\phi
{\cal{R}}-\frac{d
V}{d\phi}+\delta(y)\frac{\sqrt{-h}}{\sqrt{-g}}\frac{\delta
{\cal{L}}_{b}(\phi)}{\delta\phi}=0.
\end{equation}
Thus the original partial differential field equations have been
reduced to an ordinary differential equation.

\section{Supergravity-style solutions}
In this section we are going to generate some  special solutions of
the field equations. In this respect, one way is to introduce a
special supergravity-style potential, $V(\phi)$, as follows [21]
\begin{equation}
V(\phi)=\frac{1}{8}\Big(\frac{dW}{d\phi}\Big)^{2}-\frac{\kappa_{5}^{2}}{6}W^{2}.
\end{equation}
Assuming $k=0$, the field equations (44) and (48) are satisfied if
\begin{equation}
{\cal{F}}=\frac{\kappa_{5}^{4}}{36}W^{2},
\end{equation}
$$
\frac{d^{2}\phi}{da^{2}}=-3\frac{\phi}{\kappa_{5}^{2}
a^{2}}+\frac{3}{2}\frac{\left(\frac
{dW}{d\phi}\right)\left(\frac{d\phi}{da}\right)}{\kappa_{5}^{2}\xi W
\phi\,a}-\frac{3}{2}\frac{\phi\left(\frac
{dW}{d\phi}\right)\left(\frac{d\phi}{da}\right)}{\kappa_{5}^{2}Wa}+\frac{1}{4}\frac
{\left(\frac{d\phi}{da}\right)^{2}}{\xi\,\phi}-{ \frac
{{\left(\frac{d\phi}{da}\right)}^{2}}{\phi}}
$$
\begin{equation}
-3\frac{\left(\frac{d\phi}{da}\right)}{a}-\frac{\left(\frac{d\phi}{da}\right)^{2}\left(\frac
{dW}{d\phi}\right)}{W}-a^{-3}+\frac{9}{4}\frac{\left(\frac
{dW}{d\phi}\right)^{2}}{\xi\kappa_{5}^{2}W^{2}\phi\,{a}^{2}}\equiv
{\cal{Q}}
\end{equation}
Equation (51) is a generalization of the result obtained in Refs.
[15,21]. This equation helps us to find the necessary condition for
the consistency of the jump conditions. From equation (51) we deduce
the following equation
\begin{equation}
\frac{d\phi}{da}=\Upsilon,
\end{equation}
where $\Upsilon=\int{\cal{Q}}da$ and $\cal{Q}$ is defined in Eq.
(51). By using of Eqs. (22), (23) and (52) we find that the
consistency of jump conditions for a ${Z}_{2}$-symmetric DGP brane
is guaranteed if
$$
\frac{\delta{\cal{L}}_{b}(\phi)}{\delta\phi}=\frac{A_{0}}{3(1-\xi\phi_{0}^{2})-2\Upsilon_{0}\kappa_{5}^{2}\xi\phi_{0}
a_{0}}\Bigg\{\frac{4\xi\phi_{0}}{A_{0}}\left(3p^{(b)}-\rho^{(b)}\right)-\frac{6\xi
h_{0}\phi_{0}}{n_{0}^{2}\kappa_{5}^{2}A_{0}}(1-\xi\phi_{0}^{2})
+\frac{24\xi\phi_{0}}{n_{0}^{2}\kappa_{4}^{2}A_{0}}\left(l_{0}-m_{0}n_{0}^{2}\right)
$$

\begin{equation}
+\Upsilon_{0}\,a_{0}
\bigg[\frac{4\xi^{2}\phi_{0}^{2}h_{0}}{n_{0}^{2}A_{0}}
+\frac{3\kappa_{5}^{2}l_{0}}{n_{0}^{2}\kappa_{4}^{2}A_{0}}-\frac{\kappa_{5}^{2}}{A_{0}}\rho^{(b)}
-\frac{8\kappa_{5}^{2}\xi^{2}\phi_{0}^{2}}{(1-\xi\phi_{0}^{2})A_{0}}\Big(\rho^{(b)}+p^{(b)}\Big)
+\frac{16\kappa_{5}^{2}\xi^{2}\phi_{0}^{2}\left(l_{0}
+m_{0}n_{0}^{2}\right)}{n_{0}^{2}\kappa_{4}^{2}(1-\xi\phi_{0}^{2})A_{0}}\bigg]\Bigg\}.
\end{equation}
It should be noticed that if there is $\phi$ dependent couplings in
the standard model Lagrangian, ${\cal{L}}_{b}$ satisfies the above
condition. Equation (53) in the minimal case with $\xi=0$ simplifies
to (see [15])
\begin{equation}
\frac{\delta
{\cal{L}}_{b}(\phi)}{\delta\phi}=\Bigg(-\frac{\kappa_{5}^{2}}{6}\rho^{(b)}
+\frac{\kappa_{5}^{2}}{2\kappa_{4}^{2}}\frac{\dot{a}^{2}_{0}}{a_{0}^{2}}\Bigg)
\Bigg(-\frac{6}{\kappa_{5}^{2}W}\frac{dW}{d\phi}\Bigg)_{0}.
\end{equation}

\subsection{Exponential Superpotentials}
In this section, we consider the following exponential form of the
superpotential which is motivated by string/M-theory [21]
\begin{equation}
W=c\Bigg[\frac{e^{-{\alpha}_{1}{\phi}}}{{\alpha}_{1}}+{s}\frac{e^{{\alpha}_{2}{\phi}}}{{\alpha}_{2}}\Bigg]
,
\end{equation}
where $s=\pm1$ and $\alpha_1\geq\left|\alpha_2\right|$. For
$\alpha_2=0$, we use the following form of the superpotential
\begin{equation}
W=c\Bigg[\frac{e^{-{\alpha}_{1}{\phi}}}{{\alpha}_{1}}+\frac{s}{\kappa_{5}}\Bigg]
.
\end{equation}
The corresponding potentials obtained from (49) are
\begin{equation}
V=\frac{c^{2}}{8}\Bigg[\Bigg(1-\frac{4{\kappa_{5}}^{2}}{3{\alpha}_{1}^2}\Bigg)e^{-2{\alpha}_{1}{\phi}}+
\Bigg(1-\frac{4{\kappa_{5}}^{2}}{3{\alpha}_{2}^{2}}\Bigg)e^{2{\alpha}_{2}{\phi}}-2{s}
\Bigg(1+\frac{4{\kappa_{5}}^{2}}{3{\alpha}_1{\alpha}_{2}}\Bigg)e^{({\alpha}_{2}-{\alpha}_{1}){\phi}}\Bigg]
.
\end{equation}
and (for ${\alpha}_{2}=0$)
\begin{equation}
V=\frac{c^{2}}{8}\Bigg[\Bigg(1-\frac{4{\kappa_{5}}^{2}}{3{\alpha}_{1}^2}\Bigg)e^{-2{\alpha}_{1}{\phi}}-2{s}
\frac{4{\kappa_{5}}}{3{\alpha}_{1}}e^{-{\alpha}_{1}{\phi}}-\frac{4}{3}\Bigg].
\end{equation}
For $V$ bounded from below, only some values of the parameters are
allowed [21].\\

By using these exponential superpotentials and also equations (49)
and (51) one can deduce the evolution of the scalar field with
respect to scale factor and therefore the cosmology of the model is
obtained. Equation (51) is a complicated nonlinear second order
differential equation with no analytical solutions. So, we have
solved this equation numerically the results of which are shown in
forthcoming figures. Depending on the value of $s$ and
${\alpha}_{2}$ and also the sign of ${\alpha}_{2}$, there will be a
variety of cosmological evolution on the brane with several
interesting implications. To be more specific, in which follows we
discuss each of these choices of $s$ and $\alpha_{2}$ separately. We
note that the solutions which start at $a=0$ are interesting because
these solutions provide a big bang style cosmology.
\begin{figure}[htp]
\begin{center}\includegraphics{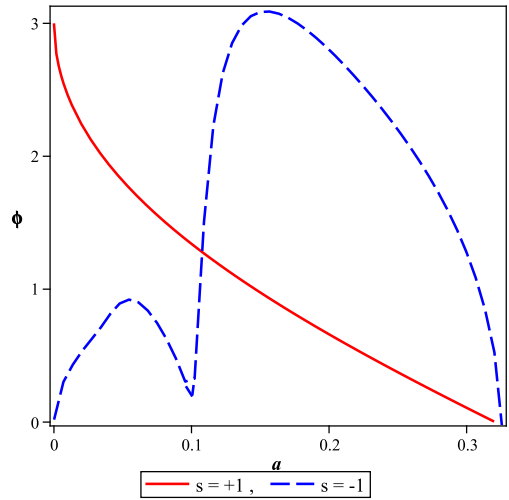} \vspace{6.5cm}
\end{center}
\caption{\small {Evolution of the scalar field with respect to the
scale factor for the case with ${\alpha}_{2}=+1$ and
$\alpha_{1}=+2$. In the case with $s=+1$ (the red-solid line), the
scalar field decreases with scale factor until in some value of $a$
it tends to zero. In the case with $s=-1$ (blue-dashed line), as
scale factor increases, the scalar field increases firstly and then
decreases. By more increment of the scale factor, $\phi$ increases
again and then decreases towards zero.}}
\end{figure}
\begin{figure}[htp]
\begin{center}\includegraphics{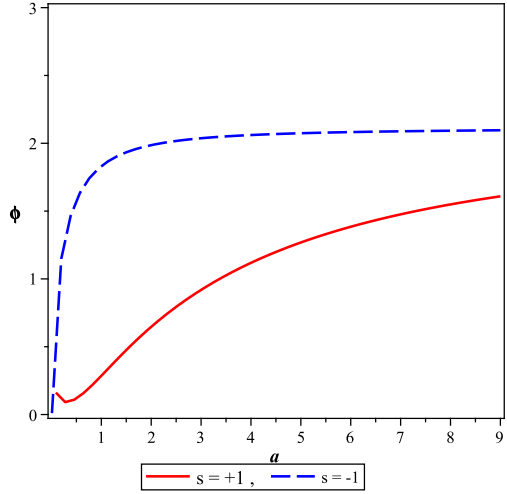} \vspace{6.5cm}
\end{center}
\caption{\small {Evolution of the scalar field versus the scale
factor for ${\alpha}_{2}=-1$ and $\alpha_{1}=+2$. In the case with
$s=+1$ (the red-solid line), the scalar field decreases firstly with
scale factor towards a minimum and then increases. In the case with
$s=-1$ (the blue-dashed line), as scale factor increases, the scalar
field increases too.}}
\end{figure}
\begin{figure}[htp]
\begin{center}\includegraphics{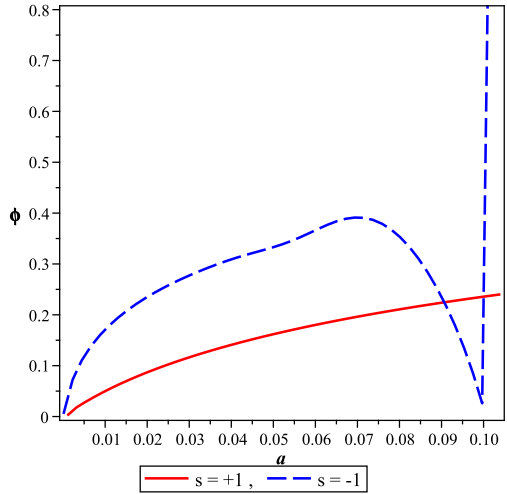} \vspace{6.5cm}
\end{center}
\caption{\small {Evolution of the scalar field versus the scale
factor for ${\alpha}_{2}=0$ and $\alpha_{1}=+2$. With $s=+1$ (the
red-solid line), the scalar field increases with scale factor. With
$s=-1$ (the blue-dashed line), as scale factor increases, the scalar
field decreases firstly and then decreases towards a minimum. After
that it increases rapidly by more increment of the scale factor.}}
\end{figure}\\

\begin{itemize}
{\item {$\alpha_{2}>0$\,:}

First, we choose $\alpha_{1}=+2$ and $\alpha_{2}=+1$. In this case
we obtain the evolution of the scalar field with respect to the
scale factor for both $s=+1$ and $s=-1$. As figure 1 shows, for
$s=+1$ the scalar field decreases by increment of the scale factor
and tends to zero at some values of the scale factor, but for $s=-1$
the behavior of the scalar field is different. For $s=-1$, at first,
the scalar field increases by the scale factor until a maximum and
then begins reduction. After that, it increases again and then
experiences another reduction until reaches zero at some values of
the scale factor. Since we expect that the scalar field to reduce by
scale factor and tend to zero finally, so it seems that these choice
of values for $\alpha_{1}$, $\alpha_{2}$ and $s$ are reliable. It
should be noted that in plotting this figure (and also forthcoming
figures) we have set $\xi=\frac{1}{6}$, $\kappa_{5}=1$, $\kappa_{4}=1$ and $c=1$.}\\

{\item {\bf$\alpha_{2}<0$\,:}

Next, we choose $\alpha_{1}=+2$ and $\alpha_{2}=-1$. In this case we
obtain the evolution of the scalar versus the scale factor for both
$s=+1$ and $s=-1$, the results of which are shown in figure 2. For
$s=+1$, the scalar field decreases firstly while the scale factor
increases and then increases until scale factor tends to infinity.
For $s=-1$, the scalar field starts at $a=0$ and increases as the
scale factor tends to infinity. So, it seems that these choice of
values for $\alpha_{1}$, $\alpha_{2}$ and $s$ cannot lead
to a cosmologically viable result. }\\

{\item {\bf$\alpha_{2}=0$\,:}

Finally, we choose $\alpha_{1}=+2$ and $\alpha_{2}=0$, so we use
equation (56) for superpotential. Figure 3 shows the behavior of the
scalar field versus the scale factor for $s=+1$ and $s=-1$. For
$s=+1$ the scalar field starts from zero at $a=0$ and increases by
scale factor. But for $s=-1$, the scalar field starts at $a=0$ and
increases by scale factor towards a maximum, then reduces by more
increment of the scale factor and finally experiences a rapid
increment again. So, these choice of values for $\alpha_{1}$,
$\alpha_{2}$ and $s$ cannot lead one to a cosmologically viable
result.}\\
\end{itemize}
We note that in the minimal case, just for $\alpha_{2}>0$ and
$s=+1$, the scalar field reduces by increment of the scale factor
and tends to zero in some value of $a$ (see [15]).

\begin{figure}[htp]
\begin{center}\includegraphics{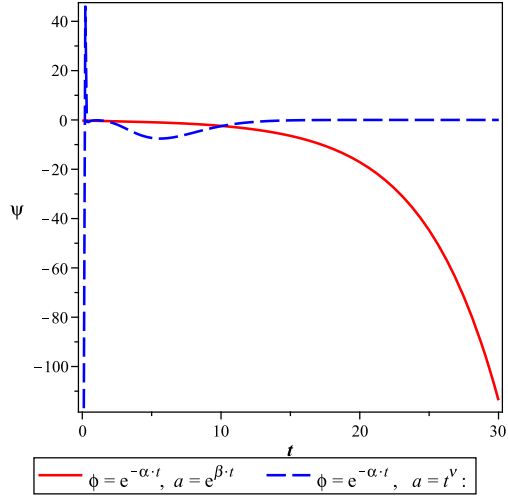} \vspace{6.5cm}
\end{center}
\caption{\small {Evolution of $\Psi$ versus the scalar field
(corresponding to $s=+1$). The behavior of this parameter clarifies
the status of the energy conservation. In the case with $a=e^{\beta
t}$ (the red-solid line), $\Psi$ is always negative. This states
that as the universe expands, the energy leaks off the brane into
the bulk. In the case with $a=t^\nu$ (the blue-dashed line), $\Psi$
is positive in some time intervals. This means that in these
intervals, the energy is sucked onto the brane by expansion of the
universe.}}
\end{figure}

\begin{figure}[htp]
\begin{center}\includegraphics{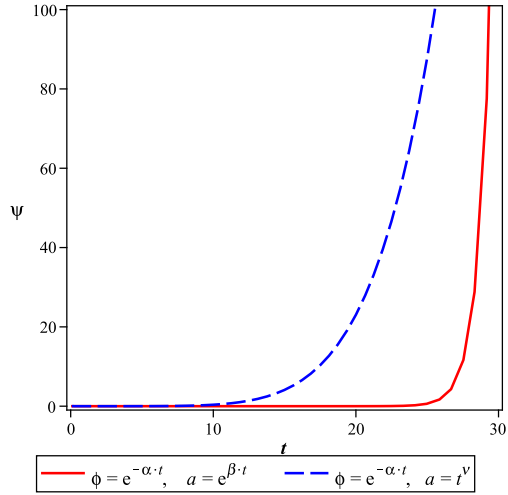} \vspace{6.5cm}
\end{center}
\caption{\small {Evolution of $\Psi$ versus the scalar field
(corresponding to $s=-1$). In both cases with $a=e^{\beta t}$ (the
red-solid line) and $a=t^\nu$ (the blue-dashed line), $\Psi$ is
always positive. This states that as the universe expands, suction
of energy onto the brane occurs.}}
\end{figure}
As we have said earlier, the presence of a bulk scalar field
(specially as in our setup with a non-minimally coupled bulk scalar
field) causes the non-conservation of the energy-momentum tensor on
the brane (that is, the bulk-brane energy-momentum exchange). Since
the right hand side of Eq. (29) shows this non-conservation, we have
performed some numerical analysis on this equation to explore some
of its physical implications. We obtain the evolution of $\Psi$ with
respect to the cosmic time on the brane in order to clarify if by
expansion of the universe, there is leakage of energy from the brane
into the bulk or energy suction occurs onto the brane from the bulk.
Note that since the case with $\alpha_{2}>0$ and $s=\pm 1$ for
superpotential are more favorable in cosmological grounds, we
perform our analysis just for the mentioned choice of parameters.
Also we consider one ansatz for the scalar field and two for scale
factor as follows
\begin{equation}
\phi=\phi_{0}\,e^{-\alpha t}\,\,,\quad \quad a=a_{0}\,e^{\beta
t}\,,\quad \quad a=a_{0}\,t^{\nu}
\end{equation}
where $\alpha$, $\beta$ and $\mu$ are positive constants. We have
considered the simplest generalization of the brane energy density
as
\begin{equation}
\rho^{(b)}=W_{0}\rho
\end{equation}
where $\rho$ is proportional to the energy density of the ordinary
matter on the brane. This generalization has its origin in the fact
that matter Lagrangian on the brane, that is
${\cal{L}}_{b}(\phi)$,\, depends on the bulk scalar field, $\phi$.
We have assumed that the ordinary matter on the brane is dust with
$\omega=0$ and $\rho=\rho_{0}\,a^{-3}$. With these assumptions, we
have performed our numerical analysis and the results are as shown
in figures 4 and 5.

\begin{itemize}
{\item {\bf$s=+1$\,:}

In the case with $s=+1$ and for $a=a_{0}\,e^{\beta t}$\,,  $\Psi$ is
negative always. So, as time passes and the universe expands, the
energy leaks off the brane. For $a=a_{0}\,t^{\nu}$, in some time
interval $\Psi$ is negative while it is positive in some other time
intervals. So, in some time interval we have leakage of energy from
the brane into the bulk and in some other time interval energy is
sucked onto the brane from the bulk.}\\

{\item {\bf$s=-1$\,:}

In the case with $s=-1$ and for both $a=a_{0}\,e^{\beta t}$ and
$a=a_{0}\,t^{\nu}$, in all times $\Psi$ is positive and this means
that, for these choices of parameters, as the universe expands there
is the energy suction onto the brane.}\\
\end{itemize}

So, it seems that the case with positive $\alpha_{2}$ and $s=+1$
with exponentially evolving scalar field and scale factor leads to a
viable cosmology.

\section{Late Time Cosmology}
In this section we study the effect of a non-minimally coupled bulk
scalar field on the late time behavior in the normal branch of a
DGP-inspired braneworld model. In this regard, we firstly rewrite
the Friedmann equation (45) (see also \textbf{Appendix A}) in some
simpler form. Since we want to study the late time behavior, so we
deal with the small scalar field regime. On the other hand, the
coupling constant, $\xi$, is small ($\xi=\frac{1}{6}$) so it is
sufficient to preserve the terms up to the order of $\xi^{2}$.
Therefore the Friedmann equation simplifies to the following form
$$
H^{2}=\frac{1}{\left(-3\kappa_{5}^{2}-16\xi^{2}\phi_{0}^{2}\kappa_{5}^{2}+3\xi\phi_{0}^{2}\kappa_{5}^{2}-
24\xi^{2}\phi_{0}^{2}\kappa_{4}^{2}\right)^{2}}
\Bigg\{384\xi^{2}\phi^{2}\kappa_{4}^{4}-48\xi^{2}\phi^{2}
\kappa_{5}^{4}m+108\kappa_{4}^{4}\xi^{2}\phi^{4}
$$
$$
+3\rho^{(b)} \kappa_{5}^{4}\kappa_{4}^{2}-72\kappa_{4}^{4}\xi\phi^
{2}+18\kappa_{4}^{4}-
6\xi\phi^{2}\kappa_{5}^{4}\rho^{(b)}\kappa_{4}^{2}+12\xi^{2}\phi^{3}\kappa_{4}^{2}g
\kappa_{5}^{4}
+40\xi^{2}\phi^{2}\rho^{(b)}\kappa_{5}^{4}\kappa_{4}^{2}-6\xi\phi\kappa_{4}^{2}g\kappa_{5}^{4}
$$
$$
+72\xi^{2}\phi^{2}\kappa_{4}^{2}m\kappa_{5}^{2}
+3\xi^{2}\phi^{4}\kappa_{5}^{4}\rho^{(b)}\kappa_{4}^{2}
+24\xi^{2}\phi^{2}\kappa_{4}^{4}\rho^{(b)}\kappa_{5}^{2}
+24\xi^{2}\phi^{2}p\kappa_{5}^{4}\kappa_{4}^{2}
$$
$$
-6\kappa_{4}^{2}\Bigg[24\xi^{2}\phi^{2}\kappa_{4}^{4}\rho\kappa_{5}^{2}-72\kappa_{4}^{4}\xi\phi^{2}
+252\kappa_{4}^{4}\xi^{2}\phi^{4}+9\kappa_{4}^{4}+72\xi^{2}\phi^{2}\kappa_{4}^{2}m\kappa_{5}^{2}
-18\xi\phi^{2}\kappa_{5}^{4}\rho\kappa_{4}^{2}
$$
$$
-6\xi\phi\kappa_{4}^{2}g\kappa_{5}^{4}+24\xi^{2}\phi^{2}p\kappa_{5}^{4}\kappa_{4}^{2}
+36\xi^{2}\phi^{3}\kappa_{4}^{2}g\kappa_{5}^{4}+104\xi^{2}\phi^{2}\rho\kappa_{5}^{4}\kappa_{4}^{2}
+45\xi^{2}\phi^{4}\kappa_{5}^{4}\rho\kappa_{4}^{2}+144\xi^{2}\phi^{2}\kappa_{4}^{2}\kappa_{5}^{2}{\cal{F}}
$$
\begin{equation}
+3\rho\kappa_{5}^{4}\kappa_{4}^{2}+9\kappa_{5}^{4}{\cal{F}}-48\xi^{2}\phi^{2}\kappa_{5}^{4}m
+288\kappa_{5}^{4}{\cal{F}}\xi^{2}\phi^{2}+135\kappa_{5}^{4}{\cal{F}}\xi^{2}\phi^{4}-
54\kappa_{5}^{4}{\cal{F}}\xi\phi^{2}\Bigg]^{\frac{1}{2}}\Bigg\}_{y=0}.
\end{equation}

Now we rewrite the Friedmann equation (61) in the form of an
effective Friedmann equation as follows
\begin{equation}
H^{2}=\frac{\kappa_{4}^{2}}{3}\Big(\rho_{m}+\rho_{eff}\Big),
\end{equation}
where $\rho_{m}$ is the standard matter energy density and
$\rho_{eff}$ is the energy density corresponding to the dark energy
sector of the model. One of the properties of phantom-like behavior
is that the effective energy density of the model grows with time
(in other words, it grows by reduction of the red-shift parameter,
$z$). We avoid to write the equation of the effective energy density
since it has a simple relation with $H^{2}$ and there is no need to
write it again. We just show the behavior of the $\rho_{eff}$ with
respect to the red-shift in figure 6. It is obvious that for both
$a=a_{0}\,e^{\beta t}$ and $a=a_{0}\,t^{\nu}$, the effective energy
density grows by cosmic expansion (decreasing $z$). It should be
noted that in plotting this figure (and also, other figures), we
have used the relation $1+z=\frac{a_{0}}{a}$.

\begin{figure}[htp]
\begin{center}\includegraphics{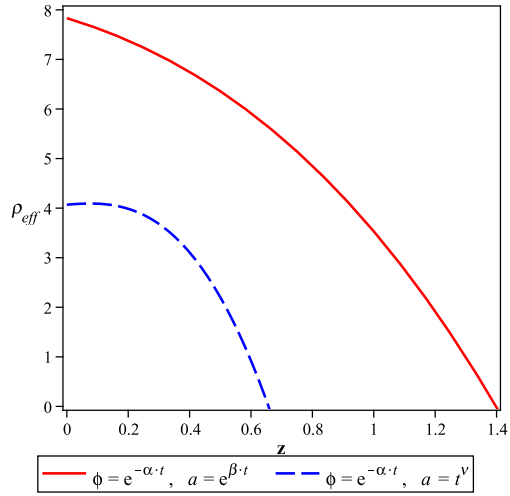} \vspace{6.5cm}
\end{center}
\caption{\small {Variation of the effective dark energy density
versus the redshift. The red-solid line is for the case with
$\phi=e^{-\alpha t}$ and $a=e^{\beta t}$. The blue-dashed line is
for the case with $\phi=e^{-\alpha t}$ and $a=t^{\nu t}$. In both
cases, the effective energy density increases when red-shift
decreases.}}
\end{figure}

\begin{figure}[htp]
\begin{center}\includegraphics{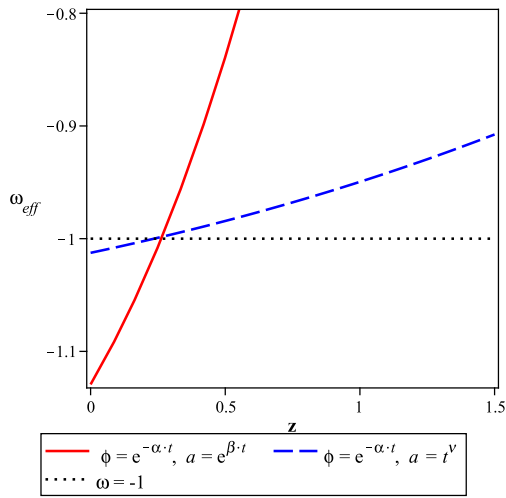} \vspace{6.5cm}
\end{center}
 \caption{\small {Variation of $\omega_{eff}$ versus the redshift. The red-solid line is for the case with
$\phi=e^{-\alpha t}$ and $a=e^{\beta t}$. The blue-dashed line is
for the case with $\phi=e^{-\alpha t}$ and $a=t^{\nu t}$. In both
cases, $\omega_{eff}$ crosses $\omega=-1$ line in $z\simeq 0.26$.}}
\end{figure}

Time evolution of the equation of state parameter gives us a
suitable background to understand the nature of dark energy. The
$\Lambda$CDM model as a candidate for dark energy has good agreement
with the recent observational data [22]. However, recent
observational data show also that the dark energy component has an
equation of state parameter $\omega<-1$ at the present epoch, while
$\omega>-1$ in the past. One way to explain these observations is to
consider a dynamical dark energy component with transient equation
of state parameter [3]. The dynamical dark energy component with
$\omega <-1$ has a \emph{phantom} nature. A cosmological model based
on the phantom fields suffers from instabilities; a phantom universe
ends up with a \emph{Big Rip} singularity. It is due to the fact
that energy density for these fields is a growing function of the
scale factor in an expanding FRW universe. It has been shown that
the normal branch of the DGP scenario has the potential to explain a
phantom-like behavior on the brane without introducing any phantom
fields neither in the bulk nor on the brane [11]. In order to
considering the evolution of the equation of state parameter in our
setup, we use the effective conservation equation as
\begin{equation}
\dot{\rho}_{eff}+3H\rho_{eff}(1+\omega_{eff})=0.
\end{equation}
\begin{figure}[htp]
\begin{center}\includegraphics{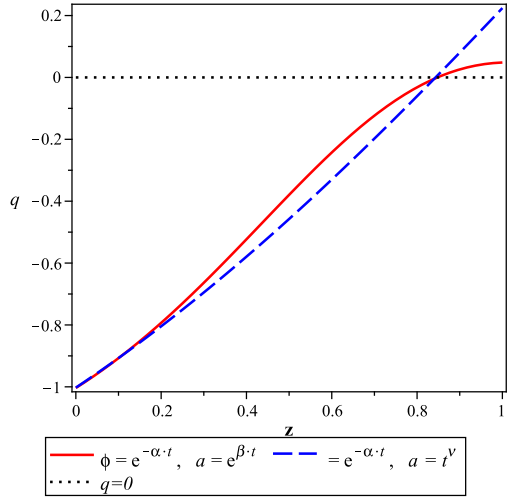} \vspace{6.5cm}
\end{center}
\caption{\small {Variation of the deceleration parameter versus the
redshift. The red-solid line is for the case with $\phi=e^{-\alpha
t}$ and $a=e^{\beta t}$. The blue-dashed line is for the case with
$\phi=e^{-\alpha t}$ and $a=t^{\nu t}$. Figure shows that the
universe, with both types of scale factors, has entered the
accelerating phase at $z\simeq 0.84$}}
\end{figure}
Note that, equation (63) is an effective conservation equation and
the effect of non-minimally coupled scalar field is hidden in the
definition of $\rho_{eff}$. By using equations (62) and (63), we
have derived the effective equation of state parameter as has been
shown in {\bf Appendix B}. The behavior of $\omega_{eff}$ with
respect to the red-shift parameter is shown in figure 7. This figure
shows that in this scenario, with both $a=a_{0}\, e^{\beta t}$ and
$a=a_{0}\,t^{\nu}$, the universe enters the phantom phase in the
near past and currently it is in the phantom phase. The transition
from quintessence to the phantom phase has occurred at $z=0.26$. So,
this model experiences a smooth crossing of the phantom divide,
$\omega_{eff}=-1$, line.

Another important parameter in cosmological evolution is the
deceleration parameter which is defined as
\begin{equation}
q=-\left[\frac{\dot{H}}{H^{2}}+1\right].
\end{equation}
A positive value of this parameter corresponds to $\ddot{a}<0$, and
this means that the universe expansion is decelerating. A negative
value of $q$ corresponding to $\ddot{a}>0$, means that the universe
expansion is positively accelerated. The calculation of the
deceleration parameter in our setup is presented in {\bf Appendix C}
and the behavior of $q$ versus $z$ is shown in figure 9. This figure
shows that the deceleration parameter becomes negative at
$z\simeq0.84$ and this means that the universe has entered into an
accelerating phase in the past at $z\simeq0.84$. So, we can say that
a warped DGP model, in the presence of a non-minimally coupled bulk
scalar field, has the phantom like behavior and can explain the late
time cosmic acceleration of the universe in an observationally
viable manner.

\section{Summary}
In this paper, we have studied the cosmological dynamics of a bulk
scalar field which is non-minimally coupled with 5D intrinsic
curvature in the DGP setup. We have derived the Bulk-brane
Einstein's equations and the scalar field's equation of motion in
this scenario. Then, by decomposing those components of the Einstein
tensor that are second derivative of the metric and matching the
resulting terms which contain Dirac delta function with the
distributional parts of the stress-energy tensor, we have derived
the jump conditions in this setup. We have found that the presence
of non-minimally coupled bulk scalar field, changes the original
jump condition of the DGP model and even changes the jump conditions
in the warped DGP model with a minimally coupled bulk scalar field.
By using the resulting jump conditions we have derived the energy
conservation equation in our setup. The non-vanishing right hand
side of this equation shows the bulk brane energy exchange in the
presence of the bulk scalar field (minimally or non-minimally
coupled with curvature). To obtain a special class of solutions for
a DGP braneworld cosmology with a bulk scalar field, by introducing
the quantity ${\cal{F}}$ as a function of $t$ and $y$, we have found
the Friedmann equation of the model and also we reduced the original
partial differential field equations to an ordinary differential
equation. We have used the superpotential method in order to
generate some solutions of the field equations of the model. By
choosing an exponential superpotential we have performed some
numerical analysis on the model parameters space. Since the
self-accelerating DGP branch has ghost instabilities, we restricted
our study to the normal DGP branch of this DGP-inspired model. We
have found that just for $\alpha_{2}>0$ and $s=\pm 1$, the scalar
field reduces by passing the time, as expected. Then, by assuming
the simplest generalization of the brane energy density
($\rho^{(b)}=W_{0} \rho$) and by choosing some ansatz for the scalar
field and the scale factor ($\phi=\phi_{0}\,e^{-\alpha t}$,
$a=a_{0}\,e^{\beta t}$ and $a=a_{0}\,t^{\nu}$), we have studied the
status of the conservation equation on the brane. Our analysis has
shown that just for $\alpha_{2}>0$ and $s=+1$, the right hand side
of the energy conservation equation is positive. This means that,
for these values of parameters the energy leaks off the brane, as
the universe expands. So, just with $\alpha_{2}>0$ and $s=+1$, the
model leads to a viable cosmology. Finally, we have studied the late
time behavior of the scenario. We have shown that the normal branch
of this DGP-inspired model, in the presence of a non-minimally
coupled bulk scalar field, realizes a phantom-like behavior, can
explain the late time cosmic acceleration and a smooth phantom
divide crossing by effective equation of state parameter on the
brane.\\

{\bf Acknowledgement}\\
This work has been supported financially by the Research Council of
the Islamic Azad University, Sari Branch, Sari, Iran.\\

{\bf Appendix A}\\
$$
H^{2}=\frac{1}{\left(-3\kappa_{5}^{2}-16\xi^{2}\phi_{0}^{2}\kappa_{5}^{2}+3\xi\phi_{0}^{2}\kappa_{5}^{2}-
24\xi^{2}\phi_{0}^{2}\kappa_{4}^{2}+24\xi^{3}\phi_{0}^{4}\kappa_{4}^{2}\right)^{2}}
\Bigg\{384\xi^{2}\phi^{2}\kappa_{4}^{4}+2048\xi^{4}\phi^{4}\kappa_{4}^{4}
$$
$$
+2048\xi^{6}\phi^{8}\kappa_{4}^{4}-72\kappa_{4}^{4}\xi^{3}\phi^{6}-4096\xi^{5}\phi^{6}\kappa_{4}^{4}-1152\xi^{3}\phi^{4}
\kappa_{4}^{4}-384\kappa_{4}^{4}\xi^{5}\phi^{8}-1152\xi^{5}\phi^{6}\kappa_{4}^{4}m
$$
$$
+48\xi^{3}\phi^{4}\kappa_{5}^{4}m-48\xi^{2}\phi^{2}
\kappa_{5}^{4}m+576\xi^{6}\phi^{8}\kappa_{4}^{4}m+576\xi^{4}\phi^
{4}\kappa_{4}^{4}m-256\xi^{4}\phi^{4}\kappa_{5}^{4}m+108\kappa_{4}^{4}\xi^{2}\phi^{4}
$$
$$
+3\rho^{(b)}
\kappa_{5}^{4}\kappa_{4}^{2}+1152\xi^{4}\phi^{6}\kappa_{4}^{4}+18\kappa_{4}^{4}\xi^{4}\phi^{8}-72\kappa_{4}^{4}\xi\phi^
{2}+18\kappa_{4}^{4}+192\xi^{4}\phi^{4}\rho^{(b)}\kappa_{5}^{2}\kappa_{4}^{4}-
6\xi\phi^{2}\kappa_{5}^{4}\rho^{(b)}\kappa_{4}^{2}
$$
$$
-144\xi^{3}{\phi}^{4}\kappa_{4}^{2}m\kappa_{5}^{2}+96\xi^{4}\phi^{5}\kappa_{4}^{4}g\kappa_{5}^{2}+
72\xi^{2}\phi^{2}\kappa_{4}^{2}m\kappa_{5}^{2}-40\xi^{3}\phi^{4}
\rho^{(b)}\kappa_{5}^{4}\kappa_{4}^{2}-48\xi^{5}\phi^{7}\kappa_{4}^{4}g\kappa_{5}^{2}
$$
$$
-48\xi^{3}\phi^{3}\kappa_{4}^{4}g\kappa_{5}^{2}+128\xi^{4}\phi^{4}p\kappa_{5}^{4}\kappa_{4}^{2}
+72\xi^{4}\phi^{6}\kappa_{4}^{2}m\kappa_{5}^{2}
-192\xi^{5}\phi^{6}\kappa_{4}^{4}\rho^{(b)}\kappa_{5}^{2}+3\xi^{2}\phi^{4}\kappa_{5}^{4}\rho^{(b)}\kappa_{4}^{2}
$$
$$
+12\xi^{2}\phi^{3}\kappa_{4}^{2}g
\kappa_{5}^{4}-32\xi^{3}\phi^{3}\kappa_{5}^{4}\kappa_{4}^{2}g-24\xi^{3}\phi^{4}p\kappa_{5}^{4}\kappa_{4}^{2}
+40\xi^{2}\phi^{2}\rho^{(b)}\kappa_{5}^{4}\kappa_{4}^{2}-6\xi\phi\kappa_{4}^{2}g\kappa_{5}^{4}
+24\xi^{4}\phi^{6}\kappa_{4}^{4}\rho^{(b)}\kappa_{5}^{2}
$$
$$
-6\xi^{3}\phi^{5}\kappa_{4}^{2}
g\kappa_{5}^{4}+192\xi^{4}\phi^{4}\kappa_{4}^{4}p\kappa_{5}^{2}-48\xi^{3}\phi^{4}\kappa_{4}^{4}\rho^{(b)}\kappa_{5}^{2}
+24\xi^{2}\phi^{2}\kappa_{4}^{4}\rho^{(b)}\kappa_{5}^{2}-192\xi^{5}\phi^{6}\kappa_{4}^{4}p\kappa_{5}^{2}
$$
$$
+128\xi^{4}\phi^{4}\rho^{(b)}\kappa^{4}\kappa_{4}^{2}+24\xi^{2}\phi^{2}p\kappa_{5}^{4}\kappa_{4}^{2}
+32\xi^{4}\phi^{5}\kappa_{4}^{2}g\kappa_{5}^{4}
$$
$$
\pm2\bigg[\kappa_{4}^{4}\left(\xi\phi^{2}-1\right)^{2}\left(32\xi^{2}\phi^{2}-3\xi\phi^{2}+3\right)^{2}
\Big(192\xi^{2}\phi^{2}\kappa_{4}^{4}+1024\xi^{4}\phi^{4}\kappa_{4}^{4}+1024\xi^{6}\phi^{8}\kappa_{4}^{4}
-36\kappa_{4}^{4}\xi^{3}\phi^{6}
$$
$$
-2048\xi^{5}\phi^{6}\kappa_{4}^{4}-576\xi^{3}\phi^{4}\kappa_{4}^{4}
-192\kappa_{4}^{4}\xi^{5}\phi^{8}
-96\kappa_{5}^{4}{\cal{F}}\xi^{3}\phi^{4}+256\xi^{4}\phi^{4}\kappa_{5}^{4}{\cal{F}}+576\xi^{6}\phi^{8}\kappa_{4}^{4}{\cal{F}}
$$
$$
+576\xi^{4}\phi^{4}\kappa_{4}^{4}{\cal{F}}-18\kappa_{5}^{4}{\cal{F}}\xi\phi^{2}+9\kappa_{5}^{4}{\cal{F}}\xi^{2}\phi^{4}
+96\kappa_{5}^{4}{\cal{F}}\xi^{2}\phi^{2}-1152\xi^{5}\phi^{6}\kappa_{4}^{4}{\cal{F}}-1152\xi^{5}\phi^{6}\kappa_{4}^{4}m
$$
$$
+48\xi^{3}\phi^{4}\kappa_{5}^{4}m-48\xi^{2}\phi^{2}\kappa_{5}^{4}m+576\xi^{6}\phi^{8}\kappa_{4}^{4}m
+576\xi^{4}\phi^{4}\kappa_{4}^{4}m
-256\xi^{4}\phi^{4}\kappa_{5}^{4}m+54\kappa_{4}^{4}\xi^{2}\phi^{4}
$$
$$
+3\rho^{(b)}
\kappa_{5}^{4}\kappa_{4}^{2}+576\xi^{4}\phi^{6}\kappa_{4}^{4}+9\kappa_{4}^{4}\xi^{4}\phi^{8}-36\kappa_{4}^{4}\xi\phi^{2}+
9\kappa_{4}^{4}+9\kappa_{5}^{4}{\cal{F}}
+192\xi^{4}\phi^{4}\rho^{(b)}\kappa_{5}^{2}\kappa_{4}^{4}-6\xi\phi^{2}\kappa_{5}^{4}\rho^{(b)}\kappa_{4}^{2}
$$
$$
-144\xi^{3}\phi^{4}\kappa_{4}^{2}m\kappa_{5}^{2}+96\xi^{4}\phi^{5}\kappa_{4}^{4}g\kappa_{5}^{2}
+72\xi^{2}\phi^{2}\kappa_{4}^{2}m\kappa_{5}^{2}-40\xi^{3}\phi^{4}\rho^{(b)}\kappa_{5}^{4}\kappa_{4}^{2}
-48\xi^{5}\phi^{7}\kappa_{4}^{4}g\kappa_{5}^{2}-48\xi^{3}\phi^{3}\kappa_{4}^{4}g\kappa_{5}^{2}
$$
$$
-288\xi^{3}\phi^{4}\kappa_{4}^{2}\kappa_{5}^{2}{\cal{F}}+144\xi^{2}\phi^{2}\kappa_{4}^{2}\kappa_{5}^{2}{\cal{F}}
+128\xi^{4}\phi^{4}p\kappa_{5}^{4}\kappa_{4}^{2}+72\xi^{4}\phi^{6}\kappa_{4}^{2}m\kappa_{5}^{2}
-192\xi^{5}\phi^{6}\kappa_{4}^{4}\rho^{(b)}\kappa_{5}^{2}
$$
$$
+3\xi^{2}\phi^{4}\kappa_{5}^{4}\rho^{(b)}\kappa_{4}^{2}-
768\kappa_{4}^{2}\xi^{5}\phi^{6}\kappa_{5}^{2}{\cal{F}}+144\xi^{4}\phi^{6}\kappa_{4}^{2}\kappa_{5}^{2}{\cal{F}}
+768\kappa_{4}^{2}\xi^{4}\phi^{4}\kappa_{5}^{2}{\cal{F}}
+12\xi^{2}\phi^{3}\kappa_{4}^{2}g\kappa^{4}
$$
$$
-32\xi^{3}\phi^{3}\kappa_{5}^{4}\kappa_{4}^{2}g-24\xi^{3}\phi^{4}p\kappa_{5}^{4}\kappa_{4}^{2}+
40\xi^{2}\phi^{2}\rho^{(b)}\kappa_{5}^{4}\kappa_{4}^{2}-6\xi\phi\kappa_{4}^{2}g\kappa_{5}^{4}
+24\xi^{4}\phi^{6}\kappa_{4}^{4}\rho^{(b)}\kappa_{5}^{2}-6\xi^{3}\phi^{5}\kappa_{4}^{2}g\kappa_{5}^{4}
$$
$$
+192\xi^{4}\phi^{4}\kappa_{4}^{4}p\kappa_{5}^{2}-48\xi^{3}\phi^{4}\kappa_{4}^{4}\rho^{(b)}\kappa_{5}^{2}
+24\xi^{2}\phi^{2}\kappa_{4}^{4}\rho^{(b)}\kappa_{5}^{2}-192\xi^{5}\phi^{6}\kappa_{4}^{4}p\kappa_{5}^{2}
+128\xi^{4}\phi^{4}\rho^{(b)}\kappa_{5}^{4}\kappa_{4}^{2}
$$
$$
+24\xi^{2}\phi^{2}p\kappa_{5}^{4}\kappa_{4}^{2}+32\xi^{4}\phi^{5}\kappa_{4}^{2}g\kappa_{5}^{4}\Big)\bigg]^{\frac{1}{2}}\Bigg\}_{0}
$$
\\\\

{\bf Appendix B}\\
$$
\omega_{eff}=-\frac{1}{3}\Bigg\{\Bigg[-96\xi^{2}\phi\kappa_{5}^{4}m
\dot{\phi}
-48\xi^{2}\phi^{2}\kappa_{5}^{4}\dot{m}+144\xi^{2}\phi\kappa_{4}^{2}m
\kappa_{5}^{2}\dot{\phi}+72\xi^{2}
\phi^{2}\kappa_{4}^{2}\dot{m}\kappa_{5}^{2}-12\xi\phi\kappa_{5}^{4}\rho^{(b)}\kappa_{4}^{2}
\dot{\phi}-\dot{\rho}_{m}
$$
$$
+48\xi^{2}\phi\kappa_{4}^{4}\rho^{(b)}\kappa_{5}^{2}\dot{\phi}+24\xi^{2}\phi^{2}{\kappa_{4}}^{4}
\dot{\rho}^{(b)}\kappa_{5}^{2}+48\xi^{2}\phi
p\kappa_{5}^{4}\kappa_{4}^{2}\dot{\phi}
+24\xi^{2}\phi^{2}\dot{p}\kappa_{5}^{4}\kappa_{4}^{2}+36\xi^{2}\phi^{2}\kappa_{4}^{2}g
\kappa_{5}^{4}\dot{\phi}
+12\xi^{2}\phi^{3}\kappa_{4}^{2}\dot{g}\kappa_{5}^{4}
 $$
$$
-6\xi\phi^{2}\kappa_{5}^{4}\dot{\rho}^{(b)}
\kappa_{4}^{2}+80\xi^{2}\phi\rho^{(b)}\kappa_{5}^{4}\kappa_{4}^{2}\dot{\phi}+40\xi^{2}\phi^{2}
\dot{\rho}^{(b)}\kappa_{5}^{4}\kappa_{4}^{2}-6\xi
\dot{\phi}\kappa_{4}^{2}g \kappa_{5}^{4}
-6\xi\phi\kappa_{4}^{2}\dot{g}\kappa_{5}^{4}+12\xi^{2}\phi^{3}\kappa_{5}^{4}\rho^{(b)}
\kappa_{4}^{2}\dot{\phi}
 $$
$$
+3\xi^{2}\phi^{4}\kappa_{5}^{4}\dot{\rho}^{(b)}\kappa_{4}^{2}
+3\dot{\rho}^{(b)}\kappa_{5}^{4}\kappa_{4}^{2}+432\kappa_{4}^{4}\xi^{2}\phi^{3}\dot{\phi}
+768\xi^{2}\phi\kappa_{4}^{4}\dot{\phi}-144\kappa_{4}^{4}\xi\phi\dot{\phi}
-3\kappa_{4}^{8}\Bigg(3\dot{\rho}^{(b)}\kappa_{5}^{4}\kappa_{4}^{2}+48\xi^{2}\phi
\kappa_{4}^{4}\rho^{(b)}\kappa_{5}^{2}\dot{\phi}$$
$$
+48\xi^{2}\phi
p\kappa_{5}^{4}\kappa_{4}^{2}\dot{\phi}+108\xi^{2}\phi^{2}\kappa_{4}^{2}g
\kappa_{5}^{4}\dot{\phi}+208\xi^{2}\phi\rho^{(b)}
\kappa_{5}^{4}\kappa_{4}^{2}\dot{\phi}+180\xi^{2}\phi^{3}\kappa_{5}^{4}\rho^{(b)}
\kappa_{4}^{2}\dot{\phi}+288\xi^{2}\phi\kappa_{4}^{2}\kappa_{5}^{2}{\cal{F}}\dot{\phi}
+9\kappa_{5}^{4}\dot{{\cal{F}}}
$$
$$
-12\xi\phi\kappa_{4}^{2}g\kappa_{5}^{4}\dot{\phi}-96\xi^{2}\phi
\kappa_{5}^{4}m\dot{\phi}+72\xi^{2}\phi^{2}\kappa_{4}^{2}\dot{m}\kappa_{5}^{2}-18\xi\phi^{2}\kappa_{5}^{4}
\dot{\rho}^{(b)}\kappa_{4}^{2}
+24\xi^{2}\phi^{2}\kappa_{4}^{4}\dot{\rho}^{(b)}\kappa_{5}^{2}+24\xi^{2}\phi^{2}
\dot{p}\kappa_{5}^{4}\kappa_{4}^{2}
 $$
$$
+36\xi^{2}\phi^{3}\kappa_{4}^{2}\dot{g}\kappa_{5}^{4}
+104\xi^{2}\phi^{2}\dot{\rho}^{(b)}\kappa_{5}^{4}\kappa_{4}^{2}+45\xi^{2}\phi^{4}\kappa_{5}^{4}
\dot{\rho}^{(b)}\kappa_{4}^{2}-6\xi\phi^{2}{\kappa_{4}}^{2}\dot{g}\kappa_{5}^{4}
+144\xi^{2}\phi^{2}\kappa_{4}^{2}\kappa_{5}^{2}\dot{{\cal{F}}}+576\kappa_{5}^{4}{\cal{F}}
\xi^{2}\phi \dot{\phi}
$$
$$
+540\kappa_{5}^{4}{\cal{F}}\xi^{2}
\phi^{3}\dot{\phi}-108\kappa_{5}^{4}{\cal{F}}
\xi\phi\dot{\phi}-36\xi\phi\kappa_{5}^{4}\rho^{(b)}
\kappa_{4}^{2}\dot{\phi}+144\xi^{2}\phi\kappa_{4}^{2}m
\kappa_{5}^{2}\dot{\phi} -48\xi^{2}\phi^{2}\kappa_{5}^{4}\dot{m}
+1008\kappa_{4}^{4}\xi^{2}\phi^{3}\dot{\phi}
$$
$$
-144\kappa_{4}^{4}\xi\phi\dot{\phi}+288\kappa_{5}^{4}\dot{{\cal{F}}}\xi^{2}\phi^{2}+135\kappa_{5}^{4}\dot{{\cal{F}}}
\xi^{2} \phi^{4}-54\kappa_{5}^{4}\dot{{\cal{F}}}\xi\phi^{2}\Bigg)
\Bigg(24\xi^{2} \phi^{2}\kappa_{4}^{4}\rho^{(b)}
\kappa_{5}^{2}-72\kappa_{4}^{4}\xi \phi^{2}+9\kappa_{4}^{4}
 $$
$$
+252\kappa_{4}^{4}\xi^{2}\phi^{4}+72\xi^{2}\phi^{2}\kappa_{4}^{2}m\kappa_{5}^{2}-18\xi\phi^{2}\kappa_{5}^{4}\rho^{(b)}
\kappa_{4}^{2}-6\xi \phi^{2}\kappa_{4}^{2}g
\kappa_{5}^{4}+24\xi^{2}\phi^{2}p\kappa_{5}^{4}\kappa_{4}^{2}+36\xi^{2}\phi^{3}\kappa_{4}^{2}g
\kappa_{5}^{4}
 $$
$$
+104\xi^{2}\phi^{2}\rho^{(b)}\kappa_{5}^{4}\kappa_{4}^{2}+45\xi^{2}
\phi^{4}\kappa_{5}^{4}\rho^{(b)}\kappa_{4}^{2}+144\xi^{2}\phi^{2}\kappa_{4}^{2}\kappa_{5}^{2}{\cal{F}}
+3\rho^{(b)}\kappa_{5}^{4}\kappa_{4}^{2} +9\kappa_{5}^{4}{\cal{F}}
-48\xi^{2}\phi^{2}\kappa_{5}^{4}m
 +288\kappa_{5}^{4}{\cal{F}}\xi^{2}
\phi^{2}
$$
$$
+135\kappa_{5}^{4}{\cal{F}}
\xi^{2}\phi^{4}-54\kappa_{5}^{4}{\cal{F}} \xi\phi^{2}\Bigg)^{-1/2}
\Bigg]\bigg(3\kappa_{5}^{2}+16\xi^{2}\phi^{2}\kappa_{5}^{2}-3\xi\phi^{2}\kappa_{5}^{2}+24\xi^{2}\phi^{2}\kappa_{4}^{2}
\bigg)^{-4}H^{-3}
$$
$$
-2H^{-1}\bigg(3\kappa_{5}^{2}+16\xi^{2}\phi^{2}\kappa_{5}^{2}-3\xi\phi^{2}\kappa_{5}^{2}
+24\xi^{2}\phi^{2}\kappa_{4}^{2} \bigg)^{-1}
\bigg(32\xi^{2}\phi\kappa_{5}^{2}\dot{\phi}-6\xi\phi\kappa_{5}^{2}\dot{\phi}
+48\xi^{2}\phi\kappa_{4}^{2}\dot{\phi}\bigg)\Bigg\}_{0}-1
$$
\\\\

{\bf Appendix C}\\
$$
q=-1-\frac{1}{2H}\Bigg\{\Bigg[768\kappa_{4}^{4}\xi^{2}\phi\dot{\phi}-144\kappa_{4}^{4}\xi\phi\dot{\phi}
+432\kappa_{4}^{4}\xi^{2}\phi^{3}\dot{\phi}-96\xi^{2}\phi\kappa_{5}^{4}m\dot{\phi}
-48\xi^{2}\phi^{2}\kappa_{5}^{4}\dot{m}-6\xi\phi\kappa_{4}^{2}\dot{g}\kappa_{5}^{4}
$$
$$
-12\xi\phi\kappa_{5}^{4}\rho\kappa_{4}^{2}\dot{\phi}-6\xi{\phi}^{2}\kappa_{5}^{4}\dot{\rho}^{(b)}\kappa_{4}^{2}
-6\xi\dot{\phi}\kappa_{4}^{2}g\kappa_{5}^{4}+80\xi^{2}\phi\rho\kappa_{5}^{4}\kappa_{4}^{2}\dot{\phi}
+40\xi^{2}\phi^{2}\dot{\rho}^{(b)}\kappa_{5}^{4}\kappa_{4}^{2}+48\kappa_{4}^{4}\xi^{2}\phi\rho\kappa_{5}^{2}\dot{\phi}
$$
$$
+24\kappa_{4}^{4}\xi^{2}\phi^{2}\dot{\rho}^{(b)}\kappa_{5}^{2}+144\xi^{2}\phi^{2}\kappa_{4}^{2}m\kappa_{5}^{2}\dot{\phi}
+72{\xi}^{2}\phi^{2}\kappa_{4}^{2}\dot{m}\kappa_{5}^{2}+48\xi^{2}\phi\,p\kappa_{5}^{4}\kappa_{4}^{2}\dot{\phi}+24
\xi^{2}\phi^{2}\dot{p}^{(b)}\kappa_{5}^{4}\kappa_{4}^{2}
$$
$$
+12\xi^{2}\phi^{3}\kappa_{5}^{4}\rho\kappa_{4}^{2}\dot{\phi}+3\xi^{2}\phi^{4}\kappa_{5}^{4}\dot{\rho}^{(b)}\kappa_{4}^{2}
+3\dot{\rho}^{(b)}\kappa_{5}^{4}\kappa_{4}^{2}-3\bigg\{9\kappa_{5}^{4}\dot{{\cal{F}}}\kappa_{4}^{4}
+3\kappa_{4}^{6}\dot{\rho}^{(b)}\kappa_{5}^{4}+104\kappa_{4}^{6}\xi^{2}\phi^{2}\dot{\rho}^{(b)}\kappa_{5}^{4}
$$
$$
+45\kappa_{4}^{6}\xi^{2}\phi^{4}\kappa_{5}^{4}\dot{\rho}^{(b)}+36\kappa_{4}^{6}
\xi^{2}\phi^{3}\dot{g}\kappa_{5}^{4}-48\kappa_{4}^{4}\xi^{2}\phi^{2}\kappa_{5}^{4}\dot{m}
-6\kappa_{4}^{6}\xi\dot{\phi}g\kappa_{5}^{4}-6
\kappa_{4}^{6}\xi\phi\dot{g}\kappa_{5}^{4}+24\kappa_{4}^{8}\xi^{2}\phi^{2}\dot{\rho}^{(b)}\kappa_{5}^{2}
$$
$$
+72\kappa_{4}^{6}\xi^{2}\phi^{2}\dot{m}
\kappa_{5}^{2}+24\kappa_{4}^{6}\xi^{2}\phi^{2}\dot{p}^{(b)}\kappa_{5}^{4}+
144\xi^{2}\phi^{2}\kappa_{5}^{2}\kappa_{4}^{6}\dot{{\cal{F}}}+288\kappa_{5}^{4}\dot{{\cal{F}}}\kappa_{4}^{4}\xi^{4}\phi^{6}
-54\kappa_{5}^{4}\dot{{\cal{F}}}\\kappa_{4}^{4}\xi\phi^{2}
$$
$$
+135\kappa_{5}^{4}\dot{{\cal{F}}}\kappa_{4}^{4}\xi^{2}
\phi^{4}+208\kappa_{4}^{6}{\xi}^{2}\phi\rho\kappa_{5}^{4}\dot{\phi}+
180\kappa_{4}^{6}\xi^{2}\phi^{3}\kappa_{5}^{4}\rho\dot{\phi}+108\kappa_{4}^{6}\xi^{2}\phi^{2}g\kappa_{5}^{4}\dot{\phi}
-96\kappa_{4}^{4}\xi^{2}\phi\kappa_{5}^{4}m\dot{\phi}
$$
$$
+48\kappa_{4}^{8}\xi^{2}\phi\rho
\kappa_{5}^{2}\dot{\phi}+144\kappa_{4}^{6}\xi^{2}\phi
m\kappa_{5}^{2}\dot{\phi}+48\kappa_{4}^{6}\xi^{2}\phi
p\kappa_{5}^{4}\dot{\phi}+288\xi^{2}\phi\kappa_{5}^{2}\kappa_{4}^{6}{\cal{F}}\dot{\phi}
+1728\kappa_{5}^{4}{\cal{F}}\kappa_{4}^{4}\xi^{4}\phi^{5}\dot{\phi}
$$
$$
-108\kappa_{5}^{4}{\cal{F}}\kappa_{4}^{4}\xi\phi\dot{\phi}
+540\kappa_{5}^{4}{\cal{F}}\kappa_{4}^{4}\xi^{2}\phi^{3}\dot{\phi}+
1008\kappa_{4}^{8}\xi^{2}\phi^{3}\dot{\phi}-144\kappa_{4}^{8}\xi\phi
\dot{\phi}+768\kappa_{4}^{8}\xi^{2}\phi\dot{\phi}\bigg\}\bigg\{\Big(9\kappa_{5}^{4}{\cal{F}}\kappa_{4}^{4}
$$
$$
+384\kappa_{4}^{8}\xi^{2}\phi^{2}+252\kappa_{4}^{8}\xi^{2}\phi^{4}-72\kappa_{4}^{8}\xi\phi^{2}
+3\kappa_{4}^{6}\rho\kappa_{5}^{4}+104\kappa_{4}^{6}\xi^{2}\phi^{2}\rho\kappa_{5}^{4}+45\kappa_{4}^{6}
\xi^{2}\phi^{4}\kappa_{5}^{4}\rho-6\kappa_{4}^{6}\xi\phi
g\kappa_{5}^{4}
$$
$$
+36\kappa_{4}^{6}\xi^{2}\phi^{3}g
\kappa_{5}^{4}+9\kappa_{4}^{8}-48\kappa_{4}^{4}\xi^{2}\phi^{2}\kappa_{5}^{4}m+24\kappa_{4}^{8}\xi^{2}\phi^{2}
\rho\kappa_{5}^{2}+72\kappa_{4}^{6}\xi^{2}\phi^{2}m\kappa_{5}^{2}+24
\kappa_{4}^{6}\xi^{2}\phi^{2}p\kappa_{5}^{4}
$$
$$
+144\xi^{2}\phi^{2}\kappa_{5}^{2}\kappa_{4}^{6}{\cal{F}}+288\kappa_{5}^{4}{\cal{F}}\kappa_{4}^{4}\xi^{4}\phi^{6}-
54\kappa_{5}^{4}{\cal{F}}\kappa_{4}^{4}\xi\phi^{2}+135\kappa_{5}^{4}{\cal{F}}\kappa_{4}^{4}
\xi^{2}\phi^{4}\Big)^{\frac{1}{2}}\bigg\}^{-1}
\Bigg]\bigg(144\xi^{2}\phi^{2}\kappa_{5}^{2}\kappa_{4}^{2}+9\kappa_{5}^{4}
$$
$$
+96\xi^{2}\phi^{2}\kappa_{5}^{4}+9\xi^{2}\phi^{4}\kappa_{5}^{4}-18\xi\phi^{2}\kappa_{5}^{4}\bigg)^{-1}
-{H}^{2}\left(288\xi^{2}\phi\kappa_{5}^{2}\kappa_{4}^{2}\dot{\phi}+192\xi^{2}\phi\kappa_{5}^{4}\dot{\phi}+36\xi^{2}
\phi^{3}\kappa_{5}^{4}\dot{\phi}-36\xi\phi\kappa_{5}^{4}\dot{\phi}\right)\Bigg\}_{0}
$$

\end{document}